\documentclass[prl,showpacs,twocolumn,superscriptaddress]{revtex4}
\usepackage{hyperref,graphicx,dcolumn}
\usepackage{amsfonts,amsmath,amssymb,bm}
\usepackage{color}

\newcommand{\epsdir}{./}

\newcommand{\myFig}[5]{ %
\begin{figure}[htb] 
\begin{center} 
\includegraphics[width=#1\columnwidth,height=#2\columnwidth,clip=true]{\epsdir/#3}
\caption{#4} \vspace{-0.5cm} \label{#5} 
\end{center} \end{figure}}

\begin{document}

\title{Magneto-mechanical interplay in spin-polarized point contacts}
\author{Maria Stamenova}
\affiliation{School of Physics, Trinity College, Dublin 2, Ireland} 
\author{Sudhakar Sahoo}
\affiliation{School of Mathematics and Physics, Queen's University of Belfast, Belfast BT7 INN, UK}
\author{Cristi\'an G. S\'anchez}
\affiliation{School of Mathematics and Physics, Queen's University of Belfast, Belfast BT7 INN, UK}
\affiliation{Unidad de Matem\'atica y F\'{\i}sica, Faculatad de Ciencias Qu\'{\i}micas, INFIQC, Universidad Nacional de C\'ordoba, Ciudad Universitaria, 5000 C\'ordoba, Argentina}
\author{Tchavdar N. Todorov}
\affiliation{School of Mathematics and Physics, Queen's University of Belfast, Belfast BT7 INN, UK}
\author{Stefano Sanvito} \email[Contact email address: ]{sanvitos@tcd.ie}
\affiliation{School of Physics, Trinity College, Dublin 2, Ireland}

\begin{abstract}
We investigate the interplay between magnetic and structural dynamics in ferromagnetic atomic point contacts. In particular, we look at the effect of the atomic relaxation on the energy barrier for magnetic domain wall migration and, reversely, at the effect of the magnetic state on the mechanical forces and structural relaxation. We observe changes of the barrier height due to the atomic relaxation up to 200\%, suggesting a very strong coupling between the structural and the magnetic degrees of freedom. The reverse interplay is weak, i.e.\! the magnetic state has little effect on the structural relaxation at equilibrium or 
under non-equilibrium, current-carrying conditions.        
\end{abstract}


\maketitle

Existing experimental techniques are capable of constructing low-dimensional magnetic constrictions consisting of a tiny number of atoms and measuring their transport properties \cite{chopra,exp1D,viret}. In this rapidly evolving field experiments and theoretical simulations are closely entangled and the latter can provide important information on the fundamental dynamics at the atomic scale. The modeling of atomic-sized ferromagnetic devices under bias requires the combined description of electron transport and of the local magnetization and structural dynamics at the atomic level \cite{stepanyuk}. In fact, since the spin-polarized current transfers both spin and charge, it exerts a twofold effect on the current-carrying structure, interlinking its structural and magnetic degrees of freedom. 

In order to investigate the mutual interplay between magnetic and structural properties, we have developed a computational scheme for evaluating spin-polarized currents and the associated current-induced forces and torques. On one hand, we are able to examine the energetics of magnetization dynamical processes as a function of the atomic rearrangements in magnetic point contacts under bias. On the other hand, we can ravel out the effect of the magnetic configuration itself on the structural relaxation. Our main finding is that this interplay is strong only in one direction. While the atomic rearrangements can modify drastically the point contact magneto-dynamics, the magnetic configuration has little effect on the atomic configuration.  

Our computational approach combines the non-equilibrium Green's function (NEGF) method for evaluating the charge density and the current \cite{alex,alex2} with a description of the magnetization dynamics in terms of quasistatic thermally-activated transitions between stationary configurations \cite{nie}. This scheme is general and is conceptually transferable \cite{nie} to first-principles Hamiltonians, for instance \cite{alex2}, within density functional theory. However, it is currently implemented in an empirical single-orbital tight-binding (TB) model \cite{ch_noble} with parameters chosen to simulate the mechanical properties of noble metals and the basic electronic structure of a generic magnetic metal. This has the benefit of being reasonably realistic while keeping the computational overheads to a minimum.

The magnetic and structural degrees of freedom both enter our model as classical variables. We associate a classical magnetic dipole moment (MM) and a set of Cartesian coordinates to each atom in the system. The magnetic state (MS) is defined by a set of angular coordinates $\Phi \equiv \{\phi_i\}$. These represent the angles of the MMs of the ``live" atoms (those contributing to the dynamics) with respect to a given direction. The structural state (SS) is defined by a set of spatial coordinates $\mathcal R \equiv \{\pmb {R}_i\}$ for these atoms. The interplay between MS and SS is then investigated by either fixing $\Phi$ and evolving $\mathcal R$ or by fixing $\mathcal R$ and evolving $\Phi$. In the first case we study how the MS affects the structural relaxation, and in the second how the SS modifies the magnetic dynamics.  

Our mixed quantum-classical Hamiltonian for this system reads
\begin{eqnarray} \label{fullH}
H(\Phi,\mathcal R)& = &\sum_{i,j,\sigma} [H^{TB}_{ij}(\mathcal {R}) + V^{\sigma}_{ij}(\Phi)] \, c_{i\sigma}^{\dagger} c_{j\sigma} + \nonumber \\
& & +\, \Omega(\mathcal{R})+W(\Phi)
\end{eqnarray}    
where $c_{i\sigma}^{\dagger}$ and $c_{i\sigma}$ are creation and annihilation operators for electrons with spin $\sigma$ at atomic site $i$. $H(\Phi,\mathcal R)$ is separated into several quantum and classical terms. The first term reads
\begin{equation} \label{TBHam}
H^{TB}_{ij}(\mathcal R)=\left[ \mathcal {E}_0 +V^{TB}_{i} (\mathcal R) \right] \delta_{ij} -\frac{\epsilon c}{2} \left. \left( \frac{a_f}{R_{ij}} \right)^{\!q} \right|_{j\ne i}
\end{equation}  
where $\mathcal{E}_0$ is the on-site energy for an isolated atom, $V^{TB}_{i}$ represents an empirical approximation to the second-order Coulomb energy at the $i$-th site
\begin{equation}
V^{TB}_{i}(\mathcal R)=\sum_{k} f_{ik}(\mathcal R) \Delta q_k=\sum_{k} \frac{\kappa_{el} \Delta q_k}{\sqrt{R_{ik}^2 + \kappa_{el}^2/U^2}}
\end{equation} 
where we have defined $f_{ik}(\mathcal R)$, $\kappa_{el}=e^{2}/4\pi\varepsilon_0=14.4$ eV\AA, $\Delta q_j=\sum_{\sigma} \rho_{jj}^{\sigma}-(\rho_0)_{jj}$ is the deviation of the electron occupancy of site $j$ ($\rho_{jj}^{\sigma}$ being a diagonal term of the density matrix for spin $\sigma$) from its equilibrium non-interacting value $\rho_0$. The last term in Eq. (\ref{TBHam}) is the intersite hopping integral, depending as an inverse power $q=4$ on the interatomic distance $R_{ij}=|\pmb{R}_i-\pmb{R}_j|$. Here we use $\epsilon=7.8680$ meV, $c=139.07$, $a_f=4.08$ \AA, which are parameters fitted to elastic properties of bulk gold \cite{ch_noble}. The value of the on-site Coulomb repulsion is $U=7$ eV and the band filling is $\rho_0=0.7$ electrons per atom. This non-integer band filling effectively accounts for the \textit{s-p-d} hybridization. 

The structure of our atomic point-contact system is presented in Figure \ref{contact}. It consists of a linear chain of three atoms attached to two semi-infinite leads with a simple cubic structure and $3 \times 3$ atom cross-section. The lead magnetizations are polarized along the $z$-axis and they are anti-parallel (AP) to each other. The MMs of the atoms in the chain are allowed to rotate in the $(x-z)$-plane, and are described by the set of polar coordinates $\Phi=\{\phi_i\}$ ($i=1,2,3$) about the $z$-axis. In this setup an abrupt domain wall (DW) is formed in the constriction. Note that our model does not include spatial spin anisotropy, therefore the direction of the spin-quantization axis $z$ is arbitrary. In this sense a Ne\'el and a Bloch wall are physically identical within our model.         

\myFig{1.0}{0.6}{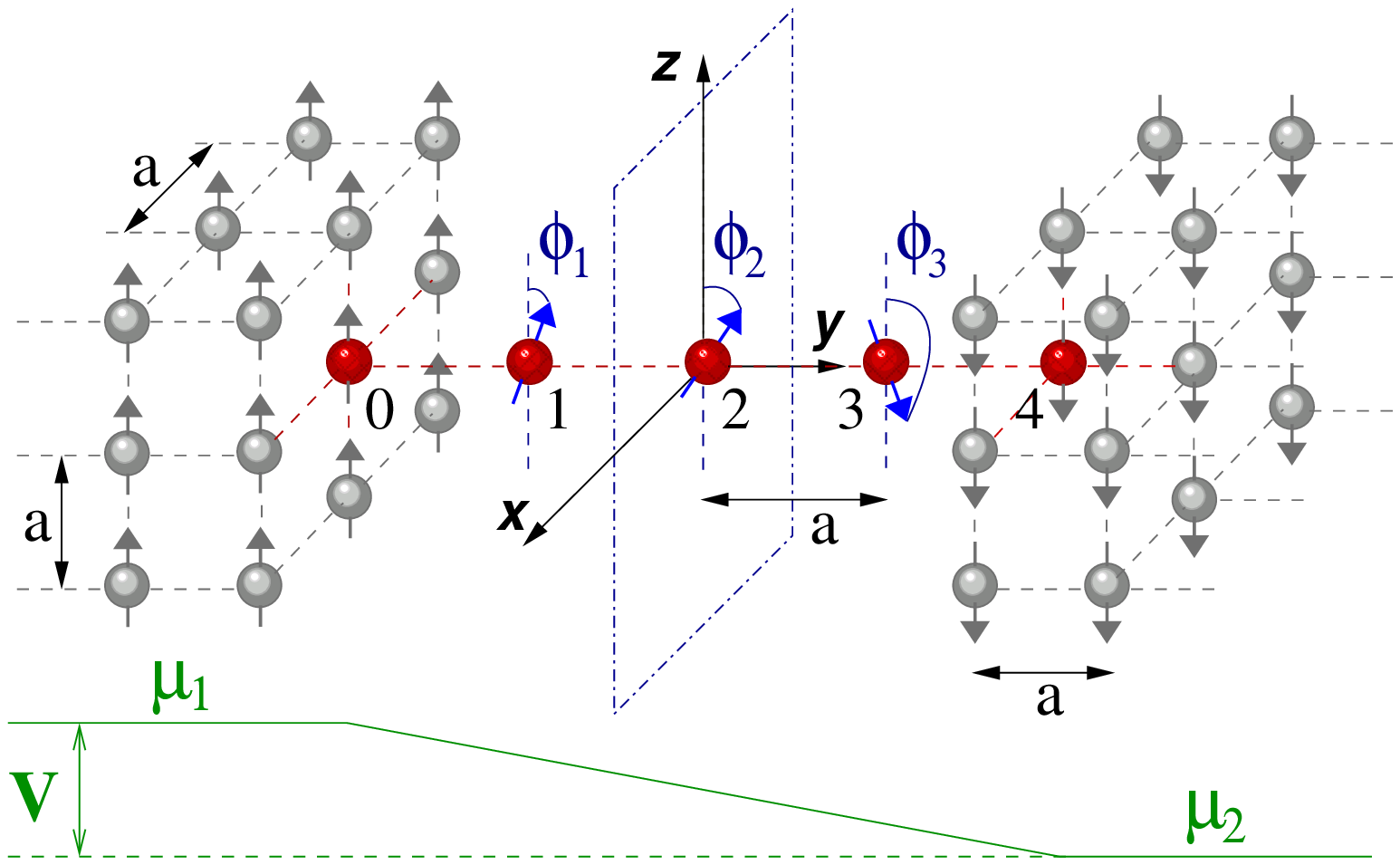}{(Color online) Schematic atomic configuration of the magnetic point contact investigated in this work. Some distances are magnified for clarity.}{contact}

The spin-dependent term in the electronic part of Eq. (\ref{fullH}) is approximated by a classical Heisenberg-type interaction as follows
\begin{equation}
V^{\sigma}_{ij}(\Phi)=-J \pmb{\sigma}\cdot \pmb{S}_i \delta_{ij}=-\sigma_z \frac{J}{2}\cos{(\phi_i)} \delta_{ij} \,, 
\end{equation}  
where $\pmb{\sigma}=(0,0,\frac{1}{2}\sigma_z)$ is the spin of the current-carrying electrons, collinear and polarized along the $z$ axis ($\sigma_z=\pm 1$) and $J>0$ is the exchange integral. We assume that the local MMs, associated with the local spins, are constant in magnitude \cite{stepanyuk}, and impose $|\pmb{S}_i|$=1.

The remaining terms in Eq. (\ref{fullH}) involve only classical variables. We have adopted a repulsive pair potential which decays as an inverse power law \cite{ch_noble}, \footnote{The pair potential and the hopping integrals
are taken to zero, between first and second nearest neighbors in the simple cubic lattice, via
a smooth tail in the range 2.8-3.3 \AA.}
\begin{equation}
\Omega(\mathcal R)=\frac{1}{2}\sum_{i,j\neq i} \Omega_{ij}(\mathcal R) = \frac{\epsilon}{2}\sum_{i,j\neq i} \left( \frac{a_f}{R_{ij}}\right)^{11}\, ,
\end{equation} 
and a Heisenberg-type nearest-neighbors spin-spin interaction between the localized spins
\begin{equation}
W(\Phi)=-\frac{J_{dd}}{2}\sum_{i,j\neq i} \pmb{S}_i \cdot \pmb{S}_j = -\frac{J_{dd}}{2}\sum_{i,j\neq i}\cos{(\phi_i-\phi_j)} \, ,
\end{equation} 
where $J_{dd}>0$ is a parameter describing the strength of the intersite exchange interaction. Typical values for the exchange parameters are $J=1$ eV and $J_{dd}=50$ meV, which are of the same order of magnitude as those for bulk ferromagnetic metals within the $s-d$ model \cite{zener}, and those derived from the Curie temperature within the Heisenberg model \cite{rushbrooke}. Clearly, the low coordination alters the values for the exchange parameters \cite{exp1D}, though this effect was not explicitly included here. However, we have repeated the calculation for $J$ from 0.8 eV upto 2 eV as well as for other force parameterizations (for Cu \cite{ch_noble}) and found no qualitative differences in the main physical results. 

Forces and torques, associated with the classical degrees of freedom, are derived from the Hellmann-Feynman theorem, generalized for non-equilibrium systems. \cite{ths, ventra} In our model the corresponding forces read\cite{ch_tdtb}
\begin{equation} \label{frcs}
F_{i}=-\! \sum_{j\neq i} \left[ 2 (\nabla_i H^{TB}_{ij}) \mathrm{Re}\! \left[ \rho_{ij} \right] + \Delta q_i \Delta q_j \nabla_i f_{ij} + \nabla_i \Omega_{ij}  \right] ,
\end{equation} 
where the live atoms $(i=0,\ldots,4)$ are those in the chain and their leftmost and rightmost neighbors in the leads $(i=0,4)$. Index $j$ runs over all the atoms in the self-consistent region, which apart from the chain includes two atomic planes from each lead (see Fig. \ref{contact}). The torques $T$ read \cite{nie} 
\begin{equation} \label{trqs}
T_{i}= -\frac{J}{2} s_i \sin\phi_i-\frac{J_{dd}}{2} \left[\sin(\phi_i-\phi_{i-1}) +\sin(\phi_{i+1}-\phi_i)\right] \, ,
\end{equation} 
where $s_i=\rho_{ii}^{\uparrow}-\rho_{ii}^{\downarrow}$ is the on-site spin-polarization $(i=1,2,3)$, $\phi_0=0$ and $\phi_4=\pi$ are considered frozen and aligned along the same direction as in the corresponding lead.

The quantum transport problem is solved by using the NEGF method \cite{alex,alex2,nie}. An analytical expression is used for the surface Green's function of the leads and a positive external bias $V$ is introduced as a rigid shift of $V/2$ of the on-site energies of the left-hand side electron reservoir and of $-V/2$ for the right-hand side with respect to their equilibrium value. The non-integer band-filling of the spin-split \textit{s} band in the leads generates an asymmetry in the transport properties of the spin-up and spin-down conduction channels upon left-to-right inversion of the system in its AP state. The relative spin-polarization $P^{\sigma} = (n^{\uparrow}-n^{\downarrow}) / (n^{\uparrow}+n^{\downarrow})$ of the density-of-states at the Fermi level, $n^{\sigma}(E_f)$, of the left lead in equilibrium is about -19\% (-17\%) for $J=1$ eV (2 eV). The net current is calculated as in Ref. \cite{nie} . The self-consistent density-matrix is used into Eqs. (\ref{frcs}) and (\ref{trqs}), and both forces and torques are determined.   

\myFig{1.0}{0.9}{figure02.eps}{(Color online) Displacements (in picometers) of the atoms in the chain: (a) from the uniform geometry at $V=0$; (b) as function of the DW migration reaction-coordinate $\phi_2$ at $V=1$ V; (c,d) as function of the bias voltage $V$. See text for details. Here $J=1$ eV, $J_{dd}=50$ meV.}{Pos}

The ``live" atoms in the constriction are relaxed at a given finite bias for a fixed MS. For symmetry reasons the atomic relaxation always results in displacements only along the longitudinal direction ($y$-axis). Thus, we denote the relaxed atomic positions at a bias voltage of $V$ and a magnetic state $\phi$ during relaxation by $y_V (\phi)$. The initial geometry, denoted by $y_{\rm uni}$, is that of equidistant atoms with a nearest-neighbor distance of $a=2.5$ \AA (see Fig. \ref{contact}). This is near the equilibrium bond length of a periodic 1D chain within our model \cite{ch_frc}. However, such a bond length produces a compressive stress in the bulk leads, as a result of which the leftmost and rightmost atoms are pushed slightly out of the leads and the whole chain at equilibrium shrinks by about 2\% [see Fig. \ref{Pos}(a)]. 

First we investigate how the MS of the constriction affects its structural relaxation. We relax the live atoms in $\{\phi^{(1)}\}=(0,0,\pi)$ or $\{\phi^{(2)}\}=(0,\pi,\pi)$ configuration, which represent two possible spatial positions of the DW inside the constriction, as well as for intermediate MSs with $\phi_2 \in [0,\pi]$ and $\phi_{1,3}$ such that $T_{1,3}=0$. The DW-motion induced displacements $\Delta y_{\phi} = y_V (\phi_2)-y_V(00\pi)$ are monotonic functions of $\phi_2$ [see Fig. \ref{Pos}(b), where $V=1V$]. They represent nearly rigid translations of the whole atomic chain in the direction of the electron flow. However, these displacements are very small and constitute about 3\% of the displacements from the uniform structure $\Delta y_{\rm uni}=y_0 (\{\phi^{(1,2)}\})-y_{\rm uni}$ [Fig. \ref{Pos}(a)]. The former depend weakly on the bias [Fig. \ref{Pos}(d)]. 

The effect of the bias voltage on the atomic relaxation $\Delta y_V = y_V (\phi) - y_0(\phi)$ for $\phi=\{\phi^{(1,2)}\}$ is demonstrated in Figure \ref{Pos}(c). The relative current-induced displacements are smooth functions of $V$ and promote a tendency towards dimerization. $ \left| \Delta y_V \right| /a < 0.7 \% $ (upto $V=2$ V) are of the same order of in magnitude to the displacements $\Delta y_{\phi}$. Interestingly, the former are almost insensitive to the MS $\{\phi^{(1,2)}\}$, which is a result of the weak bias-dependence of $\Delta y_{\{\phi^{(2)}\}}$ [see Fig. \ref{Pos}(d)]. Other MS, having multiple abrupt DWs [e.g. $(\pi,0,\pi)$], are found to produce an effect of similar magnitude on the atomic relaxation (not shown in this work). This establishes that the structural properties of a magnetic nano-device under bias are to a large extent independent of the magnetic state.

Next we address the converse, that is whether the structural relaxation affects the energy barrier for DW migration. This is calculated from the torque $T_2$ as the MM of atom ``2" is quasistatically rotated. The other two live MMs are continually relaxed so that their associated torques are kept at zero \cite{nie}. The net work for this rotation is then
\begin{equation} \label{work}
W(\phi_2)=-\int_0^{\phi_2} T_2 (\phi '_2) \,\rm{d}\phi '_2
\end{equation} 

Our calculated DW migration barriers (DWMB) for a few values of $J$ are shown in Figure \ref{barriers}. A key feature of these profiles is the tilt of the barrier at any finite bias even for a uniform atomic arrangement. This tilt results from the fact that in the absence of the special symmetry in the density of states that characterize the present simple cubic structure in the case of a half-filled band \cite{nie}, the leading contributions to current-induced forces and torques are linear in the bias. This asymmetry results in a preferential spatial localization of the DW at a given bias: under the present (positive) bias the $(0,0,\pi)$ configuration is stable, while the $(0,\pi,\pi)$ is at most metastable. The DW can then be driven back and forth in the constriction by an alternating current. That is an explicit realization of current-driven DW motion. 

\myFig{1.0}{0.9}{figure03.eps}{(Color online) DW-migration energy barriers at different bias $V$ for $J_{dd}=50$ meV and $J=1, 1.5, 2$ eV (panels from left to right). The solid (dashed) lines are for structure relaxed at $(0,0,\pi)$ [$(0,\pi,\pi)$]. $W(V,\phi_2)$ corresponds to geometry relaxed at the given bias $V$;  $\Delta W_0=W-W_0$ and $\Delta W_{\rm uni}=W-W_{\rm uni}$, where  $W_0$ refers to geometry relaxed at $V=0$ and $W_{\rm {uni}}$ to uniform geometry.}{barriers}

We have then investigated the contribution of atomic relaxation to the DW migration barrier. The bottom panels of Figure \ref{barriers} show $\Delta W_0=W-W_0$ and $\Delta W_{\rm uni}=W-W_{\rm {uni}}$ as functions of $\phi_2$ at different bias voltages $V$, where $W(\phi_2)$ is the DW migration work (see Eq. (\ref{work})) for atomic structure relaxed at the given bias and MS, while $W_0 (\phi_2)$ and $W_{\rm uni}(\phi_2)$ correspond to atomic structure relaxed at $V=0$ and to a uniform (unrelaxed) structure respectively. The small current-induced atomic displacements (see Fig. \ref{Pos}) systematically increase the DWMB height ($\Delta W_0>0$) by a few percent. $\Delta W_0$ does not show a dependence on the MS. 

However, the relaxation at $V=0$ from the uniform arrangement, which shortens the interatomic distance between the live MMs by about 4\% [Fig. \ref{Pos} (a)], reduces the height of DWMB approximately by $\Delta W_{\rm uni}/W (\phi_2)=25-30\%$ (see Fig. \ref{barriers}) for $J=1.5, 2$ eV. This effect can increase dramatically (to about 200\%) for exchange parameters close to magnetic phase transitions. \cite{nie}, \footnote{This occurs for instance when $J$ and $J_{dd}$ are such that the zero-bias magnetic energy landscape over the ``reaction coordinate" ($\phi_2$) changes from having two stable MSs with colinear local MM to only one stable MS with $\phi_2=\pi/2$. For $J_{dd}=50$ meV and this parameterization this occurs for $J$ between 0.8 eV and 0.9 eV.} Thus the magnetic properties are strongly affected by the geometry of the contact, especially in the region of parameters where the $J$ coupling mechanism starts competing with the direct exchange mechanism.  

\myFig{1.0}{0.9}{figure04.eps}{(Color online) $I-V$ curves for a geometry relaxed at $V$. Solid (dashed) lines represent $(0,0,\pi)$ $[(0,\pi,\pi)]$ state. Top inset: dependence of net current for $V=1$ V on the DW-migration reaction coordinate $\phi_2$, circles represent structure, relaxed at $\phi_2$ [see Fig. \ref{Pos}(b)]. Bottom inset: $\Delta I=I-I_{0/\rm{uni}}$, where $I_{0}$ ($I_{\rm uni}$) refer to relaxed at $V=0$ (uniform) structure.}{IVs}

Finally, the effect of the structural relaxation on the conductivity of the system is found to be small (Fig. \ref{IVs}). The current is clearly insensitive to the DW migration within the constriction. The overall variation of the net current for the rotation of $\phi_2$ for fixed geometry, which is in itself a small quantity, is further substantially compensated by structural rearrangement, i.e. the structure is found to respond to the magnetic rearrangement by structural adjustment, which minimizes the variation in the conductivity (see inset of Fig. \ref{IVs}). We have also found a decrease in conductivity due to relaxation of the structure from the uniform geometry. This agrees qualitatively with the findings in Ref. \cite{lmto}, although the effect we observe is much smaller in magnitude.

In conclusion we have developed a method to investigate the interplay between the magnetic and structural degrees of freedom of a ferromagnetic atomic point contact under bias. We have used it to assess the effects of the structural relaxation on the magnetic DWMB and reversely the effect of the magnetic configuration on the structural relaxation. Our main finding is that the interplay is only in one direction, that is the structural relaxation strongly modifies the DW migration barrier.

In particular, we have found that the DWMB shows a substantial asymmetry, which increases with the external bias even for a spatially symmetric system. That opens the possibility of voltage-controlled DW motion in such systems. The current-induced displacements ($\Delta y_{V} /a<0.7\%$) from the relaxed at $V=0$ structure produce a small positive shift in the DWMB height ($\sim 3\%$). This is small compared to the effect of the relaxation from the initial uniform atomic configuration at the given bias. The latter is $\Delta y_{\rm uni} /a< 4\%$ and has a much more dramatic effect on the DWMB profile, reducing the barrier height by upto 2/3 or even making the alternative MS unstable, i.e. blocking the DW migration, for some exchange parameters. That is a signiture of the strong non-linear dependence of  spin-polarized transport properties on structural rearrangements. However, structure is not affected by the DW migration under bias. This is so because interatomic forces depend on the total charge density of the current-carrying electrons, not on their spin polarization.

This work has been sponsored by the Irish Higher Educational Authority under the North South Programme for Collaborative Research. The authors wish to acknowledge the SFI/HEA Irish Centre for High-End Computing (ICHEC) for the provision of computational facilities and support.

\end{document}